\documentclass[twocolumn]{article}
\usepackage{arxiv}

\usepackage[acronym,nomain]{glossaries}
\glsdisablehyper
\newacronym{rnn}{RNN}{recurrent neural network}
\newacronym{bptt}{BPTT}{backpropagation through time}
\newacronym{fire}{FIRE}{fast inertial relaxation engine}
\newacronym{dos}{DOS}{density of states}

\makeglossaries
\usepackage{hyperref}
\usepackage{cleveref}
\usepackage{dsfont}
\usepackage{float}
\usepackage{bm}
\usepackage{subfig}
\usepackage{pifont}
\usepackage{dutchcal}
\usepackage[normalem]{ulem}
\usepackage{multirow}
\usepackage{amsmath}
\usepackage{amssymb}
\usepackage{natbib}

\newcommand{\appendixspringnetworks}{Appendix A.1}
\newcommand{\appendixerrorandoptimization}{Appendix A.2}
\newcommand{\appendixtwosteprelaxationnonlinear}{Appendix A.3}
\newcommand{\appendixtwosteprelaxationlinear}{Appendix A.4}
\newcommand{\appendixlyapunov}{Appendix A.5}
\newcommand{\appendixsteadystate}{Appendix A.6}
\newcommand{\appendixerrormemoryspan}{Appendix B.1}
\newcommand{\appendixmodecontribution}{Appendix B.2}
\newcommand{\appendixlearningrate}{Appendix B.3}

\usepackage[percent]{overpic}

\newcommand{\mybox}[1]{\textbf{\Large{#1}}}
\newcommand{\mysmallbox}[1]{\textbf{#1}}
\newcommand\D{\mathrm d}
\newcommand\E{\mathrm e}
\newcommand\J{\mathrm j}
\newcommand\norm[1]{\ensuremath{\left\lVert#1\right\rVert}}

\renewcommand{\vector}[1]{\overline{\bm{#1}}}

\newcommand{\phantomsubfloat}[1]{
        {
        \captionsetup[subfigure]{labelformat=empty}
        \subfloat[][]{#1}
    }%
}

\title{Designing precise dynamical steady states in disordered networks}
\author{\href{https://orcid.org/0000-0002-3821-1933}{\includegraphics[scale=0.06]{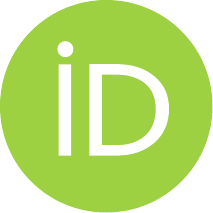}\hspace{1mm}Marc Berneman}\\
Faculty of Mechanical Engineering\\
Technion Israel Institute of Technology\\
Haifa 3200003\\
\texttt{marcberneman@campus.technion.ac.il}
\And
\href{https://orcid.org/0000-0002-7241-0741}{\includegraphics[scale=0.06]{orcid.pdf}\hspace{1mm}Daniel Hexner} \\
Faculty of Mechanical Engineering\\
Technion Israel Institute of Technology\\
Haifa 3200003\\
\texttt{danielhe@me.technion.ac.il}
}

\begin{document}
    \newgeometry{textwidth=7.25in, textheight=9.5in}

    \twocolumn[\begin{@twocolumnfalse}
                   \maketitle
                   \begin{abstract}
                       Elastic structures can be designed to exhibit precise, complex, and exotic functions.
                       While recent work has focused on the quasistatic limit governed by force balance, the mechanics at a finite driving rate are governed by Newton's equations.
                       The goal of this work is to study the feasibility, constraints, and implications of creating disordered structures with exotic properties in the dynamic regime.
                       The dynamical regime offers responses that cannot be realized in quasistatics, such as responses at an arbitrary phase, frequency-selective responses, and history-dependent responses.
                       We employ backpropagation through time and gradient descent to design spatially specific steady states in disordered spring networks.
                       We find that a broad range of steady states can be achieved with small alterations to the structure, operating both at small and large amplitudes.
                       We study the effect of varying the damping, which interpolates between the underdamped and the overdamped regime, as well as the amplitude, frequency, and phase.
                       We show that convergence depends on several competing effects, including chaos, large relaxation times, a gradient bias due to finite time simulations, and strong attenuation.
                       By studying the eigenmodes of the linearized system, we show that the systems adapt very specifically to the task they were trained to perform.
                       Our work demonstrates that within physical bounds, a broad array of exotic behaviors in the dynamic regime can be obtained, allowing for a richer range of possible applications.
                   \end{abstract}


    \end{@twocolumnfalse}]
    The elastic response of materials and structures stems from the interactions of numerous coupled degrees of freedom~\citep{lubensky2015phonons}.
    Tuning the interactions allows for harnessing the large design space, exponential in system size, to attain nontrivial responses~\citep{goodrich2015principle,coulais2016combinatorial,bhaumik2022loss,sirote2024emergent}.
    Examples of responses and functions include, signal transmission~\citep{mitchell2016strain,rocks2017designing,yan2017architecture,berry2022mechanical,pashine2023reprogrammable}, computations~\citep{kwakernaak2023counting}, and even data classification~\citep{stern2020supervised,stern2021supervised,lee2022mechanical}. To date, most work has focused on the quasistatic limit where the system is actuated slowly, transitioning between equilibrium states.

    In contrast, the dynamical regime is governed by Newton's equations.
    This regime is important since it describes actuation at a finite rate, and enables responses prohibited in quasistatics, such as responses at an arbitrary phase, responses that depend on the history and rate of applied inputs, and frequency-selective responses~\citep{jensen2003phononic,florescu2009designer,ronellenfitsch2019inverse}. Manipulating dynamics adds another layer of difficulty to the problem since it requires predicting how altering a parameter affects future trajectories.

    In this paper, we study the feasibility, constraints, and implications of designing dynamical steady states.
    Designed steady states could allow for mechanical devices whose response at long times is independent of the precise initial conditions~\citep{hermans2014automated,mandal2024learning}.
    As a test bed, we consider a single source that is actuated sinusoidally with the goal of instilling a desired periodic motion on a target site.
    We optimize using gradient descent, where the gradient is computed using automatic differentiation~\citep{werbos1994roots,bryson2018applied,rumelhart1986learning,goodrich2021designing,zu2024designing,hermans2014automated} and show that a broad array of responses can be attained.

    A central challenge in training for dynamical steady states is that the gradients of the loss function must be computed with respect to the steady state.
    However, since we rely on integrating the equations of motion, derivatives can be only computed with respect to changes to the microscopic parameters at a past `relaxation time', known as the \emph{memory span}.
    The required memory span depends on the physical parameters, most notably the amplitude and the damping coefficient.

    We characterize the phase diagram in terms of the amplitude and the damping coefficient.
    Several different competing effects bound the regime that can be trained.
    At large damping, motion is attenuated.
    At small damping and surprisingly at small amplitude, the required memory span diverges, while at large amplitude there is a transition to chaos.
    The multiple competing effects are also manifested in the convergence of the error, with large convergence times and transitions.

    The expressivity of mechanical and neural networks is often understood in terms of constraint counting.
    Within linear response, the sinusoidal motion corresponds to two integral constraints.
    This implies that a moderate number of degrees of freedom are needed in this regime.
    At a finite amplitude, we argue that there is an infinite set of constraints.
    These are associated with the harmonics of the driving frequency that are generated due to nonlinear interactions.
    Nonetheless, satisfactory convergence can be attained since high-order harmonics have a small contribution.
    However, the extra constraints give rise to slow convergence.

    We conclude by exploring how the acquired function is encoded in the normal modes.
    The spectrum changes in a manner that depends on the prescribed function, especially for small damping.
    In this limit, a small number of modes contribute.
    Depending on the phase of the response, the \gls{dos} develops either a delta function peak or valley.
    We find multiple routes by which the normal modes adapt to realize the desired motion.

    Altogether, our work provides insight into designing time-dependent responses, physical constraints, and mechanisms by which the system adapts. While the additional temporal dimension introduces new challenges, it opens the door to behaviors that by definition are time-dependent.
    Advancements in design could have relevance to metamaterials~\citep{czajkowski2024duality}, MEMS devices, micromachines, and robotics~\citep{hauser2023leveraging}.

    \subsection*{Model}
\begin{figure}[!h]
        \centering
        \subfloat{
            \begin{overpic}{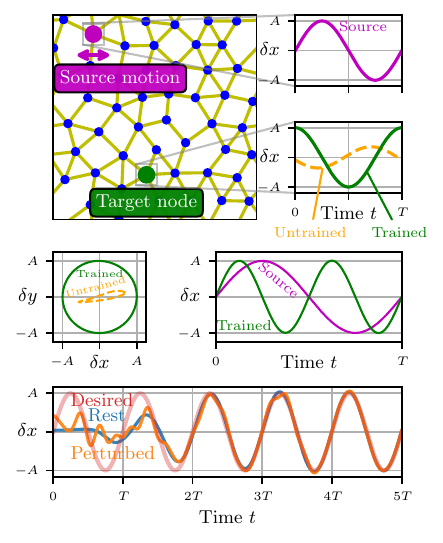}
                \put(4,92){\mybox{A}}
                \put(4,54.5){\mybox{B}}
                \put(33,54.5){\mybox{C}}
                \put(4,30){\mybox{D}}
            \end{overpic}
            \label{fig:spring_network}
        }\\
        \phantomsubfloat{\label{fig:trajectory_circular}}
        \phantomsubfloat{\label{fig:trajectory_freq_double}}
        \phantomsubfloat{\label{fig:robustness_to_restart}}
        \vspace{-2\baselineskip}
        \caption{Disordered spring network and examples of trajectories that can be attained. $\delta x$ and $\delta y$ denote the displacement the nodes.
            (A)~Example of a disordered spring network. This network is optimized to perform $\frac{\pi}{2}$ out-of-phase motion between the source and the target, with the target moving in the $x$-direction.
            (B)~Circular motion, shown in the $xy$-plane.
            (C)~Frequency doubling.
            (D)~Robustness to varying the initial conditions of the trained system. Target trajectory is shown for a trained system starting from rest, without and with a perturbation to the entire system. \\
            Parameters: (A-D) $\omega=0.5$ (source), $\gamma=0.1$, $\#\mathrm{epochs}\times\eta=10^4$. (B-D) Optimized using the nonlinear method with $n_s=10$. (A) Optimized using the linear method. $\phi = \frac{\pi}{2}$. (B,C) $A=0.2$. (D) $A=0.01$.}
        \label{fig:possible_trajectories}
    \end{figure}
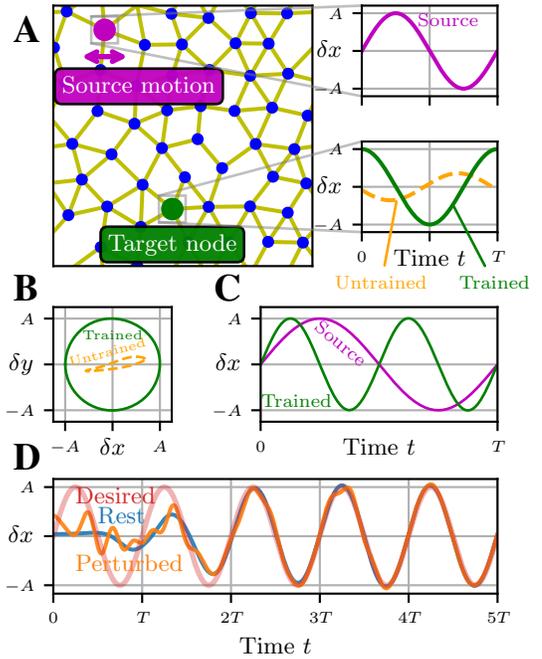

    We consider a disordered network of linear springs, which is a common model for materials~\citep{ashcroft1976solid,thorpe1990elastic,curtin1990brittle} (see \appendixspringnetworks).
    An example of a spring network is shown in \cref{fig:spring_network}.

    The potential energy of the entire network is composed of the sum of the contributions of all the linear springs,
    \begin{equation}
        V = \sum_j \frac{1}{2} k_j \left( l_j -l_{j0} \right)^2,
        \label{eq:potential_energy}
    \end{equation}
    with the sum taken over all springs $j$, $k_j$ being their stiffnesses, $l_j$ being their lengths and $l_{j0}$ being their rest lengths.

    Quasistatics concerns itself with the minimization of \cref{eq:potential_energy} with respect to the positions of all the particles.
    In this work, we shall consider the dynamics of the system.
    For every particle $i$ in our system, we can write Newton's equations of motion:
    \begin{equation}
        m_i \vector{a}_i = - \bm{\overline{\nabla}}_i V - \gamma \vector{v}_i.
        \label{eq:newton_with_friction}
    \end{equation}
    Here, the gradient of $V$ is taken with respect to the coordinates of particle $i$. A damping term is included which is proportional to the velocity $\vector{v}_i$ of the particle, $\gamma$ being the damping.
    Varying $\gamma$ allows to interpolate between the inertial regime (small $\gamma$) and the overdamped regime (large $\gamma$).
    The quasistatic limit corresponds to large $\gamma$ and slow actuation (small frequency).

    We consider driven steady states, where a \emph{source} site is periodically displaced with the goal of controlling the motion of a \emph{target} site.
    Both the source and target are chosen randomly, but care is taken to ensure that they are not connected directly by a bond.
    We assume that the source is displaced sinusoidally in the $x$-direction:
    \begin{equation}
        \begin{aligned}
            &\delta x_S (t) = A \cos (\omega t),\\
            &\delta y_S (t) = 0.
        \end{aligned}
    \end{equation}

    The desired trajectory of the target $\vector{r}_{\mathrm{target}, \mathrm{desired}}(t)$ may be sinusoidal or any other nonlinear periodic motion in the $xy$-plane.
    When the trajectory is sinusoidal with frequency $\omega$, we refer to the phase difference between source and target as $\phi$.
    In \cref{fig:possible_trajectories}, we show examples of trajectories we have trained successfully: $x_\mathrm{target} (t)$ varies as a sine with $\phi = \frac{\pi}{2}$ (\cref{fig:spring_network}), circular motion in the $xy$ plane (\cref{fig:trajectory_circular}) and lastly, sinusoidal motion at double the frequency (\cref{fig:trajectory_freq_double}).
    We note that quasistatic motion is constrained to $\phi \in \{0, \pi\}$ and that period doubling is inherently a nonlinear motion. \Cref{fig:robustness_to_restart} shows that after training, convergence is robust to varying the initial conditions.

    \subsection*{Training method} We define a cost function $\mathcal{E}$ that assigns a penalty when the motion is different from the desired motion:
    \begin{equation}
        \mathcal{E}^{(\infty)} = \frac{1}{T} \int_{0}^{T} \norm{ \vector r^{ (\infty)}_\mathrm{target} (t)- \vector r_{\mathrm{target},\mathrm{desired}} (t)}^2 \D t,
        \label{eq:error_non_normalized}
    \end{equation}
    where $T = 2 \pi / \omega$ is the period, and $\infty$ indicates that the error is evaluated in steady state.
    The error has contributions that scale differently with the amplitude $A$. In practice, we employ the normalized error $\mathcal{E}_\mathrm{norm}$, which eliminates the amplitude dependence by rescaling the different terms separately (see \appendixerrorandoptimization).

    In this paper, we optimize with respect to the rest lengths $l_{j0}$, using gradient descent:
    \begin{equation}
        \Delta l_{j0} = -\eta \frac{\partial \mathcal{E}_\mathrm{norm}^{(\infty)}}{\partial l_{j0}},
        \label{eq:update_rule}
    \end{equation}
    with $\eta$ being the learning rate.

    While in principle the gradient must be evaluated in steady-state, integration of the equations of motion provides finite time trajectories.
    When working with finite time trajectories, one must specify the time at which the rest lengths were altered.
    Imagine we begin in a steady state $\vector r^{(\infty)}(t;l_{j0})$, and alter the rest length at a past time $t_0$.
    This results in transient motion, $\vector r(t;l_{j0}+\D l_{j0})=\vector r^{(\infty)}(t;l_{j0}+\D l_{j0})+ \vector r^{(\mathrm{transient})}(t)$.
    Since both the transient and the change to the steady state scale as $\D l_{j0}$, there is a bias in the gradient, which does not vanish even when $\D l_{j0}\rightarrow 0$.

    We define the \emph{memory span} $n_s$ as the number of periods between altering the rest lengths and measuring its effect on the trajectory of the target, i.e.\ $t_0 = -\left(n_s-1 \right) T$.
    To eliminate the transient, $n_s$ should be taken to be as large as possible.
    In the field of machine learning (in particular \glspl{rnn}) this concept is also known as the unroll length~\citep{metz2021gradients}, the length of the memory buffers~\citep{mohajerin2019multistep}, or the batch time~\citep{hermans2014automated}.
    We will demonstrate that the memory span plays an important role.

    Computing the gradient requires calculating the change in trajectories as a result of altering the rest lengths at a past time.
    To this end, we employ the backpropagation algorithm, originally developed for the efficient computation of gradients in neural networks.
    In the context of \glspl{rnn}, which is most similar to our problem, the algorithm is known as \acrlong{bptt}~\citep{robinson1987utility,werbos1988generalization,mozer2013focused}.
    At each epoch $\tau$, we first compute the gradient by integrating Newton's equations of motion over $n_s$ periods.
    Then, we update the rest lengths in proportion to the gradient (\cref{eq:update_rule}).

    We also employ a direct approach that is valid in the linear regime, where the desired target motion is at the same frequency as the source.
    In the limit of small amplitudes, the full nonlinear equations can be approximated by linear equations.
    The steady state in these equations can be solved without integrating the equations of motion.
    The normalized error can therefore be obtained directly as well.
    To compute the gradient in this case, we employ automatic differentiation~\citep{werbos1994roots,bryson2018applied,rumelhart1986learning,goodrich2021designing,zu2024designing}. Comparing the nonlinear method based on \acrlong{bptt} to the direct linear method allows us to study the influence of the memory span $n_s$ and the amplitude $A$.

    All our computations are aided by JAX~\citep{jax}, JAX M.D.~\citep{jaxmd}, and JAXopt~\citep{jaxopt}.
    In addition, we use the velocity Verlet algorithm to simulate the dynamics~\citep{velocity_Verlet}.
    The \acrshort{fire} algorithm is used to relax the system to force balance~\citep{FIRE}.

    \subsection*{Linear theory \& constraint counting}
    We linearize the system and transform \cref{eq:newton_with_friction} into the frequency domain:
    \begin{equation}
        -M \omega^2 \vector X + \J \gamma \omega \vector X + K \vector X = 0 ,
        \label{eq:linearized}
    \end{equation}
    where $\vector X$ is a column vector containing the $x$ and $y$ displacements of the nodes, $K$ is the stiffness matrix such that $\Delta V \approx \frac12 \vector X^T K \vector X$ , $M = m I$ is the mass matrix, and $\J^2 = -1$. Note that we make a distinction between unconstrained degrees of freedom (free) and constrained degrees of freedom, such as the driven source. The stiffness matrix can be decomposed into free and constrained submatrices:
    \begin{equation}
        K=\begin{pmatrix}
              K_{ff} & K_{fs} \\ K_{21} & K_{22}
        \end{pmatrix},
    \end{equation}
    where $K_{ff}$ couples the free degrees of freedom, and $K_{fs}$ couples the source to the free degrees of freedom.
    The response of the target is then given by
    \begin{equation}
        X_{\mathrm{target}} = \frac{1}{m} \sum_\lambda \underbrace{\left(\left(K_{fs} \vector X_s\right) \cdot \vector u_\lambda\right) u_{\lambda, \mathrm{target}}}_{c_\lambda} f_\omega(\omega_\lambda),
        \label{eq:target_response}
    \end{equation}
    where the sum is taken over the eigenvalue-eigenvector pairs $\left(m \omega_\lambda^2, \vector u_\lambda\right)$ of $K_{ff}$, $\vector X_s$ is the motion of the source, and the weighing function is given by
    \begin{equation}
        f_\omega(\omega_\lambda) = \frac{1}{\omega^2 - \omega_\lambda^2 - \J (\gamma / m) \omega}.
        \label{eq:weighing_function}
    \end{equation}
    In \cref{eq:target_response}, $c_\lambda$ is referred to as the \emph{IO coupling} since it encodes the coupling of the source and target through the eigenvectors.

    We are seeking solutions where the motion of the target is the same amplitude as the source but possibly a different phase. In particular, we focus on two cases: in-phase ($\phi=0$), and $\frac{\pi}{2}$ out-of-phase motion ($\phi=\frac{\pi}{2}$). In terms of $X_{\mathrm{target}}$ this corresponds to,
    \begin{equation}
        \begin{aligned}
            \phi = 0: \quad & \mathfrak{Re}\{X_{\mathrm{target}}\} = A, & \mathfrak{Im}\{X_{\mathrm{target}}\} = 0, \\
            \phi = \frac{\pi}{2}: \quad & \mathfrak{Re}\{X_{\mathrm{target}}\} = 0, & \mathfrak{Im}\{X_{\mathrm{target}}\} = A.
        \end{aligned}
    \end{equation}
    These amount to two equations the system must satisfy.\footnote{Two additional equations specify the motion in y-direction}
    Since there are many degrees of freedom (number of bonds) we expect a large space of solutions.
    In the nonlinear regime, we argue that there is an infinite set of constraints.
    Driving at a finite amplitude generates harmonics at multiples of the driving frequencies.
    Each harmonic must be eliminated which introduces additional constraints.

    \section*{Results}
    As noted above, we focus on trajectories where both the source and the desired motion of the target are sinusoidal with the same frequency $\omega$ and amplitude $A$.
    We consider both a phase difference of $\phi=0$ and $\phi = \frac{\pi}{2}$.

    \subsection*{Convergence}
\begin{figure}
        \centering
        \subfloat{
            \begin{overpic}{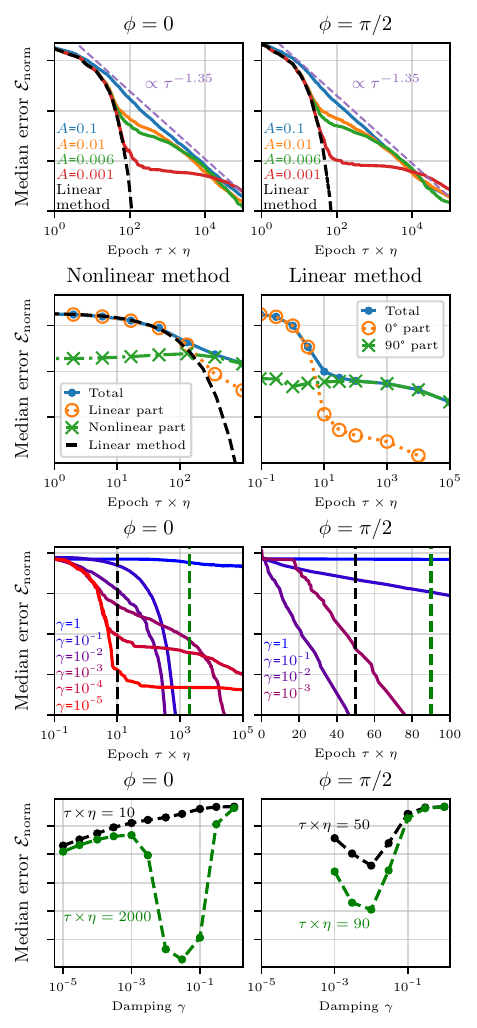}
                \put(2,96.5){\mybox{A}}
                \put(24.5,96.5){\mybox{B}}
                \put(2,72){\mybox{C}}
                \put(24.5,72){\mybox{D}}
                \put(2,47.5){\mybox{E}}
                \put(24.5,47.5){\mybox{F}}
                \put(2,23){\mybox{G}}
                \put(24.5,23){\mybox{H}}
            \end{overpic}
            \label{fig:error_amplitude_phi1}
        }\\
        \phantomsubfloat{\label{fig:error_amplitude_phi2}}
        \phantomsubfloat{\label{fig:two_step_relaxation1}}
        \phantomsubfloat{\label{fig:two_step_relaxation2}}
        \phantomsubfloat{\label{fig:error_gamma_phi1}}
        \phantomsubfloat{\label{fig:error_gamma_phi2}}
        \phantomsubfloat{\label{fig:error_gamma_phi3}}
        \phantomsubfloat{\label{fig:error_gamma_phi4}}
        \vspace{-2\baselineskip}
        \caption{Characterizing the convergence of the linear and nonlinear method. (A-F)~Median normalized error as a function of the number of epochs. (A,B)~Optimized using the nonlinear method for $\phi=0$ (A) and $\phi = \frac{\pi}{2}$~(B), for different $A$.
            (C)~Optimized using the nonlinear method, also showing the linear and nonlinear parts of the normalized error and the normalized error obtained with the linear method.
            (D)~Optimized using the linear method, also showing the in-phase and $\frac{\pi}{2}$ out-of-phase parts that comprise the total error.
            (E,F)~Optimized using the linear method for $\phi=0$ (E) and $\phi = \frac{\pi}{2}$ (F), for different $\gamma$.
            (G) and (H) show the error after a certain number of epochs (vertical dashed lines in (E) and (F)) as a function of $\gamma$, for the data in (E) and (F) respectively.\\
            Parameters: (A-H) $\omega=0.5$. (A,B) $\gamma=0.03$, $n_s = 50$. (C) $A=0.1$, $\gamma = 0.1$, $\phi = \frac{\pi}{2}$, $n_s = 10$. (D) $\gamma=0.0001$, $\phi = 0$. (F) $\gamma \geq 10^{-3}$ in (F) because of the high computational cost to simulate systems with low damping (see \appendixlearningrate).
        }
    \end{figure}
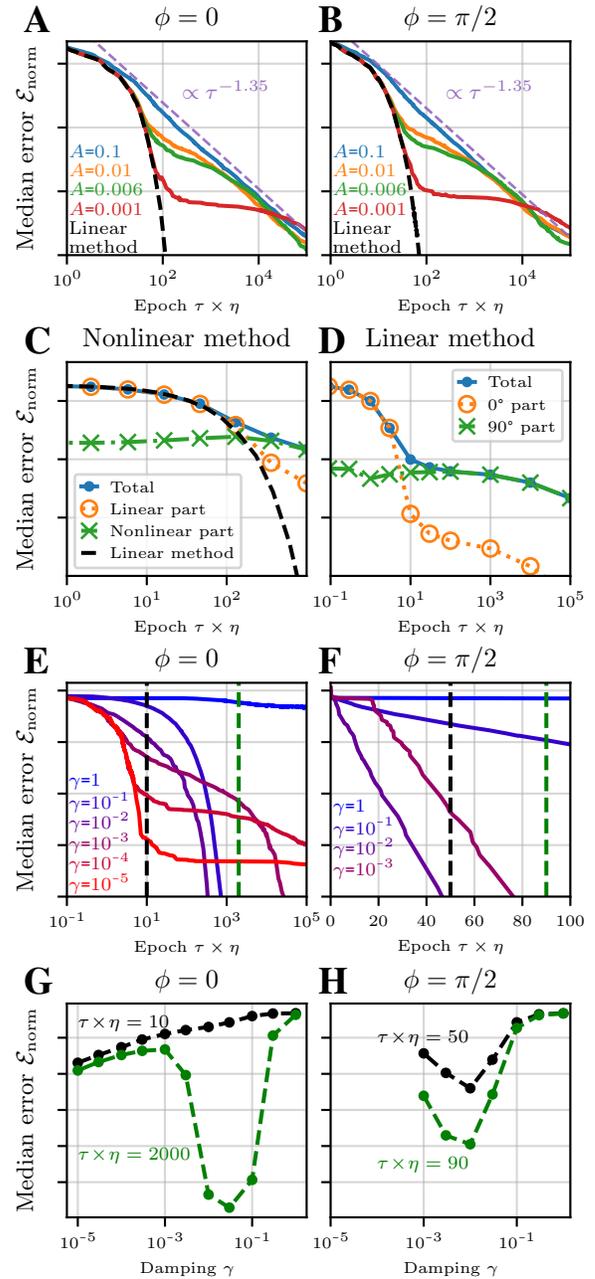

    We begin by characterizing the decrease of the error as a function of the amplitude $A$.
    In \cref{fig:error_amplitude_phi1,fig:error_amplitude_phi2}, the convergence of the normalized error is shown for $\phi=0$ and $\phi=\frac{\pi}{2}$ respectively, for different amplitudes $A$, in addition to the linear (direct) method.
    Convergence is fastest for the linear method and becomes slower with increasing amplitude.
    At early times, the error of the nonlinear curves follows the linear curve and then departs at a time that depends on the amplitude, leading to a behavior reminiscent of two-step relaxation~\citep{cavagna2009supercooled}.
    Interestingly, while the linear curves decrease exponentially, at large times all the nonlinear curves appear to follow the same power-law $\tau^{-\approx 1.35}$. Power-law convergence in gradient descent implies that the spectrum of the Hessian of the loss function is ungapped~\citep{chong2013introduction}.

    The slow convergence at a finite amplitude indicates that manipulating the nonlinear regime is challenging. As noted, in the nonlinear regime, the harmonics that are generated amount to extra constraints.
    This can be verified empirically by separating the error $\mathcal{E}_\mathrm{norm}$ into a linear part and a nonlinear part. We decompose the motion of the target into a Fourier series and associate the linear error with motion at the driving frequency, and the DC term.
    Higher-order harmonics are associated with the nonlinear contribution to the error.
    As shown in \cref{fig:two_step_relaxation1}, the error deviates from the error obtained with the linear method when the nonlinear part becomes dominant (see \appendixtwosteprelaxationnonlinear{}).

    Next, we characterize the dependence on the damping $\gamma$, which interpolates between underdamped and overdamped dynamics.
    We focus on the linear method to avoid the effects of a finite memory span $n_s$ and the amplitude $A$.
    \Cref{fig:error_gamma_phi1,fig:error_gamma_phi2} show the convergence of the normalized error for $\phi=0$ and $\phi = \frac{\pi}{2}$ respectively, for different values of the damping $\gamma$.
    For high damping the convergence is slow.
    The large damping attenuates the motion, preventing the target from oscillating at the same amplitude as the source.
    Lowering the damping initially leads to faster convergence, and then a slower convergence.

    \Cref{fig:error_gamma_phi3,fig:error_gamma_phi4} show the error after a fixed number of epochs as a function of $\gamma$.
    Convergence is fastest at an intermediate damping $\gamma$.
    Interestingly, there is a difference between $\phi = 0$ and $\phi = \frac{\pi}{2}$.
    While for $\phi = \frac{\pi}{2}$ convergence is exponential, for $\phi =0$ the error crosses over to a power-law at small $\gamma$.
    This could indicate a critical transition at a finite or vanishing value of $\gamma$.
    As shown in \cref{fig:two_step_relaxation2}, the two-step relaxation for $\phi=0$ stems from the fact that the in-phase part and out-of-phase part of the error decrease at different rates (see \appendixtwosteprelaxationlinear{}).
    The different convergence for $\phi=0$ and $\phi=\frac{\pi}{2}$ demonstrates that the system adapts differently.

    \subsection*{Phase diagram}The amplitude $A$ and the damping $\gamma$ interpolate between different dynamical regimes.
    Here, we map out convergence in terms of those parameters and discuss the relation to the underlying physical characteristics of the motion.
    To this end, we compute the normalized error after training as a function of the damping $\gamma$ and the amplitude $A$ for a fixed memory span $n_s$, shown in \cref{fig:error_phase_plot_20_0.1_0.0}.
    We find that the error is bounded by curves that delineate different behaviors:

    \begin{figure}[!h]
        \centering
        \subfloat{
            \begin{overpic}{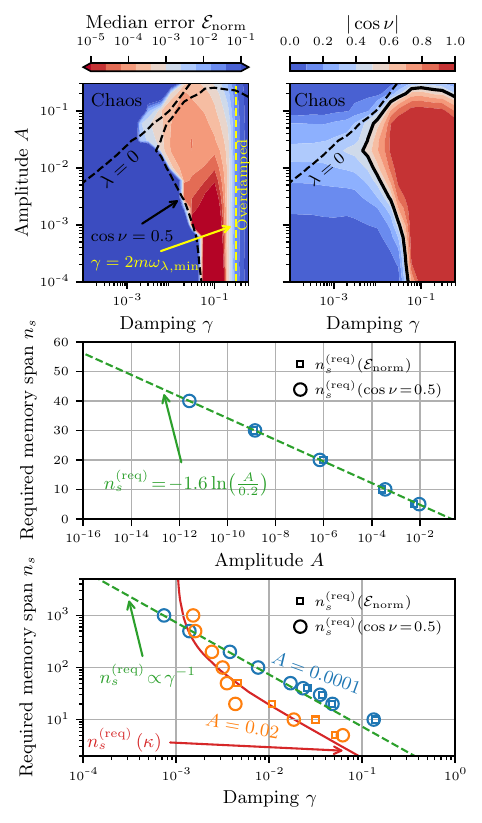}
                \put(5,96){\mybox{A}}
                \put(33,96){\mybox{B}}
                \put(5,60){\mybox{C}}
                \put(5,31){\mybox{D}}
            \end{overpic}
            \label{fig:error_phase_plot_20_0.1_0.0}
        }\\
        \phantomsubfloat{\label{fig:gradient_bias_20_0.00}}
        \phantomsubfloat{\label{fig:error_cosnu_amplitude_fit_only}}
        \phantomsubfloat{\label{fig:error_cosnu_gamma_fit_only_plus_chaos}}
        \vspace{-2\baselineskip}
        \caption{The training phase diagram and dependence on the memory span.
            (A)~Median normalized error at the end of training.
            $\cos \nu = 0.5$ is also shown, in addition to the chaos line ($\lambda = 0$), and the onset of overdamped dynamics ($\gamma = 2 m \omega_{\lambda, \min}$).
            (B)~Median cosine angle (\cref{eq:cosnu}).
            (C)~Required memory span $n_s^{(\mathrm{req})}$ delineating the regime where $\mathcal{E}_\mathrm{norm}$ converges and diverges, and the required memory span estimated by $\cos \nu(n_s^{(\mathrm{req})}) = 0.5$, both as a function of the amplitude $A$.
            (D)~Similar to (C), but with variable damping $\gamma$, shown for low and high $A$.
            For high $A$, the red filled line denotes the number of epochs needed to diminish a transient by 90\% (\cref{eq:steady-state_exp}).\\
            Parameters: (A-D) $\omega=0.5$, $\phi=0$. (A,C,D) Optimized using the nonlinear method with $\#\mathrm{epochs} \times \eta = 10^3$. (A,B) $n_s=20$. (C) $\gamma=0.1$. (D) $A=0.0001$ (low amplitude) and $A=0.02$ (high amplitude).
        }
    \end{figure}
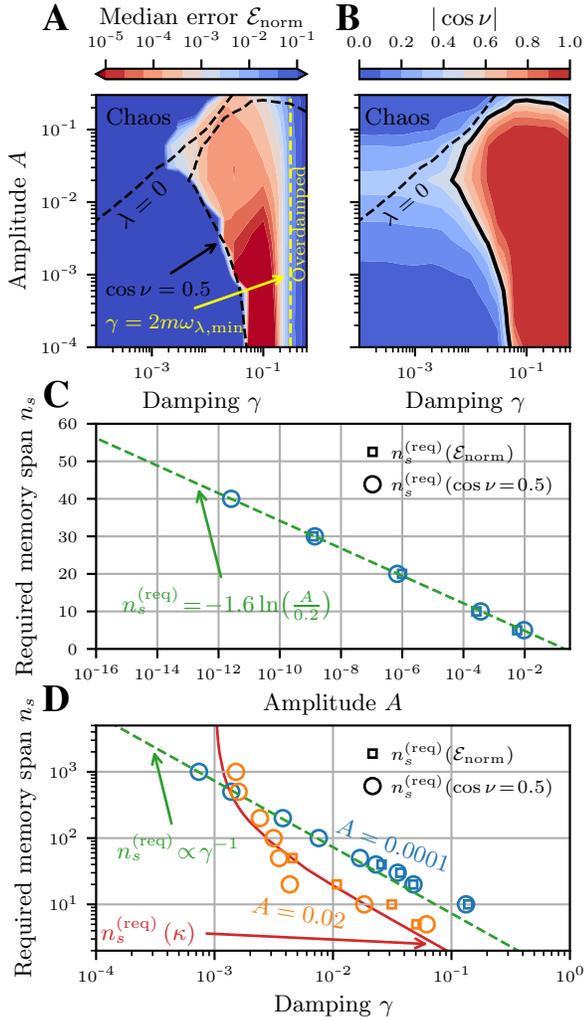

    \begin{enumerate}
        \item The onset of \emph{overdamped} dynamics, which corresponds to $\gamma = 2 m \omega_{\lambda, \min}$, where $\omega_{\lambda, \min}$ is the minimum eigenfrequency of the system.\footnote{For a single 1D spring, \cref{eq:newton_with_friction} becomes $m a = -\gamma v - k x$.
        This second order system becomes overdamped for $\gamma > 2 \sqrt{mk} = 2 m \omega_\lambda$, with $\omega_\lambda = \sqrt{k/m}$.
        For a network of springs, we require that \emph{all} modes are underdamped, and therefore the onset of overdamped dynamics is given by $\gamma = 2 m \omega_{\lambda, \min}$.}
        \item The onset of \emph{chaos}, which corresponds to $\lambda = 0$, with $\lambda$ being the Lyapunov exponent.
        The Lyapunov exponent is a measure of the rate of separation of infinitesimally close trajectories in phase space.
        Given an initial separation $\delta Z(0)$, the future separation is given approximately by $\delta Z(t) \approx \E^{\lambda t} \delta Z(0)$.
        Thus, a positive Lyapunov exponent corresponds to a chaotic system (see \appendixlyapunov{}).
        Interestingly, training can converge beyond the chaotic line, provided the memory span is sufficiently small~\citep{metz2022gradientsneed}.
        \item \emph{Gradient bias at small $\gamma$:} at small $\gamma$, the large relaxation time necessitates a large memory span $n_s$; small $n_s$ leads to a bias to the gradient which may result in catastrophic failure.
        \item \emph{Gradient bias at small $A$:} surprisingly, small amplitudes introduce a bias to the gradient, which can also be prevented by a sufficiently large memory span $n_s$.
    \end{enumerate}

    To assess the bias in the gradient, we compare the gradient computed with \acrlong{bptt} and the gradient computed with the linear method.
    The linear method is exact in the limit of vanishing amplitude $A$.
    After computing the gradients, the cosine angle is calculated:
    \begin{equation}
        \cos \nu (n_s) = \frac{\bm{\overline{\nabla}} \mathcal{E}_{\mathrm{norm},\mathrm{nonlinear}}^{(n_s)} \cdot \bm{\overline{\nabla}} \mathcal{E}_{\mathrm{norm},\mathrm{linear}}^{\phantom{(n_s)}}}{\norm{\bm{\overline{\nabla}} \mathcal{E}_{\mathrm{norm},\mathrm{nonlinear}}^{(n_s)}} \norm{\bm{\overline{\nabla}} \mathcal{E}_{\mathrm{norm},\mathrm{linear}}^{\phantom{(n_s)}}}}.
        \label{eq:cosnu}
    \end{equation}
    $\cos \nu$ is shown in \cref{fig:gradient_bias_20_0.00}.
    $\cos \nu = 1$ implies there is no bias, while a lower value implies a bias.
    The dashed line in \cref{fig:error_phase_plot_20_0.1_0.0} that marks $\cos \nu=0.5$ shows that this measure of the bias is a good indicator of convergence.
    At large $A$ this measure is less reliable due to nonlinear effects.

    \subsection*{Required memory span}
    We derive the dependence of the memory span on $\gamma$ and $A$.
    First, we assume that the system is in steady state.
    We then change the rest length of a certain spring by $\D l_0$ at time $t_0=-(n_s-1)T$.
    This induces a transient that is proportional to $\D l_0$,
    \begin{equation}
        r_\mathrm{target}^{(n_s)}(t;l_0 +\D l_0) = r_\mathrm{target}^{(\infty)}(t;l_0 +\D l_0) + g^{(n_s)}(t) \D l_0.
    \end{equation}
    The transient $g^{(n_s)}(t)$ decreases with $n_s$ and is assumed to be independent of $\D l_0$ and the amplitude.
    Using this assumption and the definition of the error in \cref{eq:error_non_normalized}, the finite time contribution to the gradient can be computed:
    \begin{equation}
        \frac{\partial \mathcal{E}^{(n_s)}}{\partial l_0} = \frac{\partial \mathcal{E}^{(\infty)}}{\partial l_0} + \frac{2}{T} \int_0^T ( r_\mathrm{target}^{(\infty)} - r_{\mathrm{target},\mathrm{desired}}) g^{(n_s)} \D t.
    \end{equation}
    The first term on the right-hand side is the steady state gradient, while the second term is the bias.
    Based on dimensional analysis and assuming linear theory: $\frac{\partial \mathcal{E}^{(\infty)}}{\partial l_0} \propto A^2$, and
    $ \int_0^T ( r_\mathrm{target}^{(\infty)} - r_{\mathrm{target},\mathrm{desired}}) g^{(n_s)} \D t \propto A \E^{-\gamma n_s T}$.
    The transient becomes important when these two terms are of the same magnitude,
    \begin{equation}
        \boxed{n_s^{(\mathrm{req})} \propto - \frac{1}{\gamma T} \ln \left (\frac{A}{A_0} \right )}.
        \label{eq:memory_span}
    \end{equation}
    Here, $A_0$ is a constant.
    This relation describes the required memory span $n_s$ so that the gradient bias can be neglected.
    As advertised, the required memory span diverges at small $\gamma$ and small $A$.

    We verify \cref{eq:memory_span} numerically with two different metrics.
    (1)~We vary $n_s$ and find the value which delineates the regime where $\mathcal{E}_\mathrm{norm}$ converges and diverges (see \appendixerrormemoryspan).
    (2)~We estimate $n_s^{(\mathrm{req})}$ with the condition $\cos \nu(n_s) = 0.5$.
    These two metrics are shown in \cref{fig:error_cosnu_amplitude_fit_only} as a function of $A$, and are approximately proportional to $- \ln \left (A / A_0 \right )$.
    In a similar fashion, \cref{fig:error_cosnu_gamma_fit_only_plus_chaos} shows that $n_s^{(\mathrm{req})}$ grows with decreasing $\gamma$, and is consistent with $\gamma^{-1}$, at least for low amplitude.

    At larger amplitudes, out of the linear regime, the relaxation rate differs from $\gamma$.
    \Cref{fig:error_cosnu_gamma_fit_only_plus_chaos} shows that $n_s^{(\mathrm{req})}$ diverges at large amplitude, which is indicative of a diverging relaxation time. We associate this divergence with the onset of chaos~\citep{strogatz2018nonlinear}.

    The convergence rate to steady-state can be estimated by comparing the change in motion for every two consecutive periods, denoted by $\delta S$ (see \appendixsteadystate). The transient approximately decays with the number of periods $n$ as,
    \begin{equation}
        \delta S(n) \approx \E^{-\kappa n} \delta S(0),
        \label{eq:steady-state_exp}
    \end{equation}
    where $\kappa$ is the convergence rate.
    The required memory span can be estimated by solving for $\delta S(n_s^{(\mathrm{req})}) = 0.1 \delta S(0)$.
    The red line in \cref{fig:error_cosnu_gamma_fit_only_plus_chaos} shows that this metric conforms to the other two metrics.

    The divergence of $n_s^{(\mathrm{req})}$  suggests that it is not possible to surpass the chaotic line.
    However, numerical experiments demonstrate that there is convergence slightly beyond the chaotic line.
    In this regime, it is beneficial to take a small memory span, since there is no true steady state in the untrained network~\citep{metz2022gradientsneed}. Note that the chaotic line is measured before training, and training may affect the occurrence of chaos.

    \section*{Nature of solutions}
    In this section, we discuss the adaptation of the system as it acquires the desired response.
    We explore how the response is encoded in the elastic properties~\citep{bhaumik2022loss,mendels2023systematic}.
    We focus on the linear regime, which is characterized by the normal modes, and their corresponding frequencies.

    \subsection*{Eigenanalysis}
\begin{figure*}
        \centering
        \hspace*{-0.3cm}
        \subfloat{
            \begin{overpic}{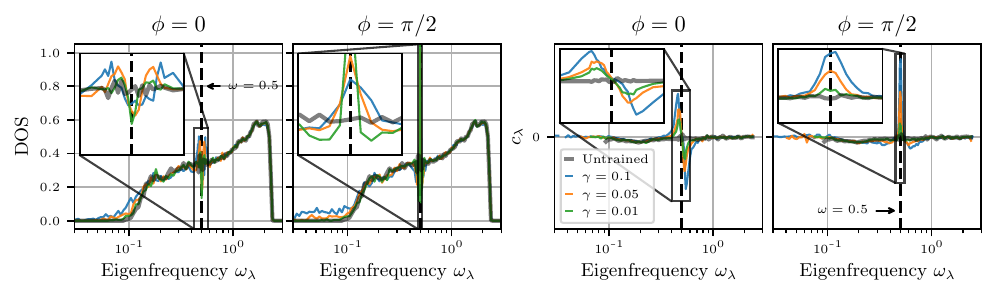}
                \put(7,26){\mybox{A}}
                \put(29,26){\mybox{B}}
                \put(55,26){\mybox{C}}
                \put(77,26){\mybox{D}}
            \end{overpic}
            \label{fig:dos1}
        }\\
        \phantomsubfloat{\label{fig:dos2}}
        \phantomsubfloat{\label{fig:couplings1}}
        \phantomsubfloat{\label{fig:couplings2}}
        \vspace{-2\baselineskip}
        \caption{Adaptation of the spring networks to the optimized task. (A-B) The \acrfull{dos} acquires a peak or a valley due to training, which becomes sharp for small $\gamma$. (C-D) IO-couplings $c_\lambda$ averaged over bins. For $\phi = 0$ (A,C) and $\phi = \frac{\pi}{2}$ (B,D) for various values of the damping $\gamma$, in addition to the untrained network.
        Parameters: optimized using the linear method with $\#\mathrm{epochs} \times \eta = 10^4$, $\omega=0.5$.}
    \end{figure*}
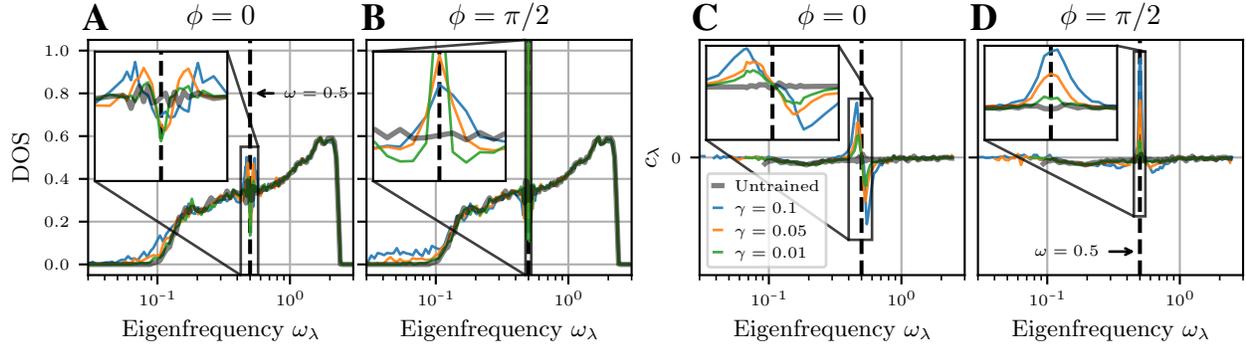

    We begin by discussing the change in the eigenfrequency spectrum, characterized by the \acrfull{dos}~\citep{liu2010jamming,zaccone2023theory}, which is shown in \cref{fig:dos1,fig:dos2} for $\phi = 0$ and $\phi = \frac{\pi}{2}$ respectively.
    We consider the effect of varying the damping $\gamma$.

    For $\phi = \frac{\pi}{2}$ (\cref{fig:dos2}), the \gls{dos} exhibits a peak at $\omega_\lambda = \omega$, which becomes narrower and taller with decreasing $\gamma$.
    The narrowing of the peak can be understood from the imaginary part of $f_\omega(\omega_\lambda)$ (see \cref{eq:weighing_function}) which also becomes increasingly peaked for decreasing $\gamma$.
    As a result, when $\gamma \rightarrow 0$, only modes at the driving frequency contribute, and therefore there must be at least a single mode precisely at $\omega$.
    Hence, the \gls{dos} develops a delta function.

    For $\phi = 0$ (\cref{fig:dos1}), a `valley' develops in the \gls{dos}, which becomes narrower and deeper as $\gamma$ decreases .
    When $\gamma \rightarrow 0$ we argue that there are no modes at $\omega$.
    This again results from the structure of the imaginary part of $f_\omega(\omega_\lambda)$, which in this limit contributes only if there is a mode at the driving frequency.
    To suppress any $\frac{\pi}{2}$ out-of-phase motion, the \gls{dos} must vanish at $\omega$.

    The emergence of the `valley' also explains the slow-down in the convergence of the error when $\gamma$ decreases, as was observed in \cref{fig:error_gamma_phi1}.
    For large $\gamma$, the $\frac{\pi}{2}$ out-of-phase motion can be suppressed by the alignment of the normal modes.
    For small $\gamma$, the $\frac{\pi}{2}$ out-of-phase motion can only be suppressed by removing any modes near the driving frequency. This, we argue, results in a transition in the mechanism by which the response is realized. Below, we argue that aligning the normal modes is easier than altering the eigenfrequencies.

    Next, we consider the individual contribution of the eigenmodes to the desired motion, given by the IO couplings $c_\lambda$ as defined in \cref{eq:target_response}.
    In \cref{fig:couplings1,fig:couplings2}, we show the IO couplings for $\phi=0$ and $\phi = \frac{\pi}{2}$ respectively.
    For $\phi=0$, the main contribution is near the driving frequency and $c_{\lambda}$ changes sign, similarly to the real part of $f_{\omega}(\omega_{\lambda})$.
    For $\phi = \frac{\pi}{2}$, the IO couplings are peaked at the driving frequency, which is similar to the imaginary part of $f_{\omega}(\omega_\lambda)$.

    The structure of the IO couplings can be understood from the linear response (see \cref{eq:target_response}).
    For $\phi=0$, the $\frac{\pi}{2}$ out-of-phase motion is suppressed, namely $\int c_{\lambda} \mathfrak{Im} \{f_\omega(\omega_\lambda)\} D(\omega_{\lambda}) \D \omega_{\lambda}= 0$.
    Since $\mathfrak{Im}\{f_\omega(\omega_\lambda)\}$ is approximately symmetric around the driving frequency, $c_{\lambda}$ is approximately anti-symmetric.
    Similarly, for $\phi=\frac{\pi}{2}$ the real contribution of the motion is suppressed by having $c_{\lambda}$ be approximately symmetric.

    Lastly, we note that $c_{\lambda}$ becomes small as $\gamma$ decreases.
    However, we note that both for the real and imaginary part of $f_{\omega} (\omega_{\lambda})$ attain larger values as $\gamma$ decreases, which compensates for the decrease in $c_{\lambda}$.

    \subsection*{Number of contributing modes}
    The motion in the linear regime can be decomposed into the eigenvectors, i.e. $\vector{X}_f = \sum_{\lambda} a_{\lambda} \vector u_\lambda$, with $a_\lambda = \vector{X}_f \cdot \vector u_\lambda$.
    Given that the eigenvectors $\vector u_\lambda$ are orthonormal, $\norm{\vector{X}_f}^2 = \sum_\lambda |a_\lambda|^2$.
    We can then define the \emph{participation ratio} as
    \begin{equation}
        P = \frac{(\sum_\lambda |a_\lambda|^2)^2}{\sum_\lambda |a_\lambda|^4},
        \label{eq:participation_ratio}
    \end{equation}
    where a value of 1 means that only a single mode is responsible for the motion of the system.
    We will use the participation ratio to quantify the number of modes that are responsible for the motion of the system.

\begin{figure}
        \centering
        \subfloat{
            \begin{overpic}{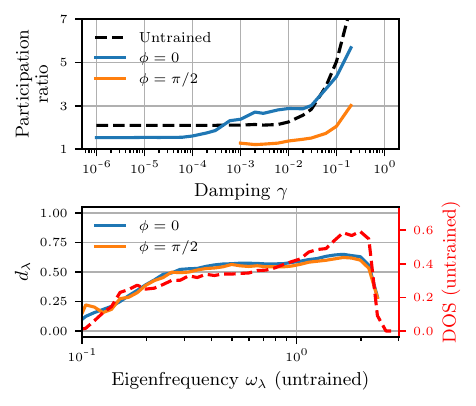}
                \put(3,81){\mybox{A}}
                \put(3,42){\mybox{B}}
            \end{overpic}
            \label{fig:dominant_mode_energy_phase}
        }\\
        \phantomsubfloat{\label{fig:average_alignment}}
        \vspace{-2\baselineskip}
        \caption{Characterizing the change to the eigenvectors and their contribution to the total motion.
            (A)~Median participation ratio (\cref{eq:participation_ratio}) as a function of $\gamma$, for $\phi = 0$ and $\phi = \frac{\pi}{2}$.
            (B)~Change in eigenvectors (\cref{eq:dlambda}), averaged over every bin, for $\phi = 0$ and $\phi = \frac{\pi}{2}$.
            In addition, the \acrfull{dos} of the untrained network is also shown.\\
            Parameters: (A,B) optimized using the linear method with $\#\mathrm{epochs} \times \eta = 10^3$, $\omega=0.5$. (A) Only showing simulations for which the median normalized error $\mathcal{E}_\mathrm{norm} \leq 10^{-3}$. (B) $\gamma = 0.1$.}
    \end{figure}

    In \cref{fig:dominant_mode_energy_phase}, we plot the median participation ratio as a function of $\gamma$ for different $\phi$.
    Both before and after optimization, the participation ratio decreases as $\gamma$ is reduced. I.e., at large $\gamma$, many modes contribute while at small $\gamma$, few contribute.
    This results from the narrowing of $|f_\omega(\omega_\lambda)|^2$ at small $\gamma$. For small enough $\gamma$, the participation ratio plateaus. The value of $\gamma$ where this occurs can be estimated by comparing the width of $|f_\omega(\omega_\lambda)|^2$ to the spacing between modes ($\mathcal{O}(1/N)$), suggesting that it is a finite-size effect.

    For $\phi = \frac{\pi}{2}$, the decrease in the participation ratio is more significant, and when $\gamma\rightarrow 0$ the motion is dominated by a single mode.
    For $\phi = 0$, the plateau appears to be different than unity, implying multiple contributing modes.

    We have looked at individual realizations for $\phi=0$ at small $\gamma$ (see \appendixmodecontribution).
    We find realizations with varying numbers of contributing modes.
    The probability of having $m$ contributing modes decreases with $m$.
    Solutions with $m>1$ do not necessarily perform worse than $m=1$.

    We note that in the limit of $\gamma \rightarrow 0$ the real and imaginary parts of $f_\omega(\omega_\lambda)$ behave differently.
    While $\mathfrak{Im}\{f_\omega(\omega_\lambda)\}$ approaches a delta function at the driving frequency, $\mathfrak{Re}\{f_\omega(\omega_\lambda)\}$ does not vanish for $\omega_{\lambda} \ne \omega$.
    This, presumably allows for multiple contributing modes when $\phi = 0$.

    \subsection*{Change in eigenvectors}
    Next, we consider the change in the eigenvectors $\vector u_\lambda$. We quantify the change in the eigenvectors using the dot product of the eigenvector before and after optimization (assuming the eigenvector has unity norm),
    \begin{equation}
    {d}
        _\lambda = 1 - |\vector u_\lambda^{(\mathrm{untrained})} \cdot \vector u_\lambda^{(\mathrm{trained})}|.
        \label{eq:dlambda}
    \end{equation}
    If the alignment does not change, ${d}_\lambda=0$. If the elements of the eigenvectors before and after optimization are uncorrelated and have zero mean, ${d}_\lambda \rightarrow 1$ with increasing system size.

    \Cref{fig:average_alignment} shows that the alignment is approximately 0.5 implying a significant change to the eigenvectors. Interestingly, the change in alignment approximately follows the \gls{dos}.
    This can be understood within perturbation theory, assuming small changes to the structure, which results in perturbations to the stiffness matrix, $\delta K_{ff}$. The lowest order change in the eigenvectors is given by~\citep{cohen1977quantum}:
    \begin{equation}
        \delta \vector u_i \approx \frac{1}{m}\sum_{j \neq i} \frac{\vector u_j^T \delta K_{ff} \vector u_i}{ \omega^2_i -  \omega^2_j} \vector u_j.
    \end{equation}
    Thus, only small changes to the structure are required to realign the eigenvectors, particularly when the density of states is dense. Note that the lowest order change in the eigenfrequencies does not have the difference in eigenfrequencies in the denominator, and therefore requires more significant changes to the structure. This is consistent, with slow relaxation at small $\gamma$ which requires changing the spectrum of the frequencies.

    \section*{Conclusion}
    We have shown that a disordered structure can be adjusted to perform specific predetermined functions. This includes responses at an arbitrary phase, as well as non-linear functions, e.g., frequency doubling.
    The advantage of acting in the dynamical regime is that a broader range of responses can be attained, and responses can be history-dependent and frequency-selective.
    Since we have optimized for steady states, this function is weakly dependent on the initial conditions.

    Training for steady states gives rise to new challenges, that are not present when optimizing with set initial conditions.
    Gradients must be computed with respect to the steady state, which is not easily accessible since the equations are nonlinear.
    Estimating the gradient with finite time trajectories obtained by integrating the equations of motion gives rise to a bias in the computed gradient.
    The bias does not self-average and can cause catastrophic failure.
    Developing methods that do not require large integration times is a central problem in the small dissipation limit.

    We have mapped out the $\gamma$-$A$ phase diagram and have shown that convergence is constrained by several different effects including chaos, large relaxation times, small amplitude bias, and strong damping.
    Interestingly, this results in a window where optimization is most successful.
    Training near chaos gives rise to diverging relaxation times.
    Nonetheless, it is possible to optimize slightly beyond the chaotic line by taking a low memory span, also noted in~\citep{metz2022gradientsneed}.

    We have also studied how the eigenmodes and eigenvalues change with training.
    Their change depends specifically on the phase difference between input and output, which can be explained through linear theory.
    In the limit of $\gamma\rightarrow 0$, the \gls{dos} develops either a delta function peak or valley.
    In this limit, the system converges even for small $\gamma$ (for the linear method), despite the diverging resonances around the individual eigenmodes.

    The linear regime presumably has a continuous space of solutions.
    Indeed, we observe multiple solutions.
    For a given phase difference there are two integral constraints: the undesired phase in the motion must vanish and the desired phase must correspond to the desired amplitude.
    These essentially amount to four constraints (including suppression of motion along the $y$-axis).
    Since there are many variable degrees of freedom (number of bonds), there should be a large number of solutions.
    The nonlinear regime appears to be more restrictive.
    There, the nonlinearities generate harmonics of the driving frequency.
    Suppressing these modes requires additional constraints, and therefore poses a greater difficulty, which is apparent in the slower convergence.

    We envision several applications.
    The structure could be used to transmit or transduce motion which could be useful for energy harvesting and energy transfer.
    Furthermore, the structure can act as a `mechanical circuit' allowing complex responses, that can sense signals, or even perform simple temporal computations.
    Unlike most machinery composed of multiple components (e.g., gears) that are individually crafted and then assembled, here a single structure can be produced using additive or subtractive manufacturing.
    Therefore, this approach could be more suitable for creating machines on a microscopic scale.

    \section*{Acknowledgments}
    I would like to acknowledge Himangsu Bhaumik and Sheng Huang for their insightful conversations.
    This work was supported by the Israel Science Foundation (grant 2385/20) and the Alon Fellowship.

    \appendix
    \section*{Appendix}

\section{Methods}
\subsection{Network preparation}
\label{appendix:spring_networks}
Spring networks are created from a packing of soft-spheres~\citep{o2003jamming}.
The particles are polydisperse, whose radii are linearly spaced.
These particles are randomly placed in a 2-dimensional deformable square box with periodic boundary conditions under constant pressure.
The pressure is calculated by using the virial equation~\citep{LOUWERSE2006138}.
The particles undergo repulsive harmonic interactions, i.e.\ when spheres overlap, the contribution to the total potential energy is
\begin{equation*}
    V_{ij} = \frac{1}{2} \left(1 - \frac{\|\vector{r}_{ij}\|}{R_i + R_j}\right)^2,
\end{equation*}
with $\|\vector{r}_{ij}\|$ being the distance between the centers of the spheres, and $R_i + R_j$ being the sum of the radii of the two spheres.
\Acrshort{fire} is then used is used to minimize the potential energy of the entire system~\citep{FIRE}.
Then, overlapping spheres are replaced with springs with unit spring constant, whose rest lengths are set to the distance between the centers of the spheres.
This way, the network is unstressed initially.
Finally, particles with less than 3 bonds are pruned, in order to avoid having dangling bonds in the network.

We generate these networks with 50 nodes before pruning and a pressure of 0.01.
This gives us networks with an average excess coordination number $\Delta Z = 0.42$, with $\Delta Z=2 N_b/N -Z_c$, where $N_b$ is the number of bonds, $N$ the number of nodes and $Z_c=2$ in the limit of large systems~\citep{maxwell1864calculation,liu2010jamming}.

We randomly assign a source, target, and 2 fixed nodes.
Care is taken to make sure that a bond doesn't connect these nodes.
We take the masses $m_i = 1$, and the spring stiffnesses $k_j = 1$.

\subsection{Error and optimization}
\label{appendix:error}
The error is defined as follows.
\begin{equation*}
    \mathcal{E} = \frac{1}{T} \int_{0}^{T} \norm{ \vector{\epsilon}(t) }^2 \D t.
\end{equation*}
with $\vector{\epsilon}(t) = \vector r^{(n_s)}_\mathrm{target} (t)- \vector r_{\mathrm{target},\mathrm{desired}} (t)$, $T$ being the period of the driving, and $n_s$ being the memory span.

We separate $\vector{\epsilon}(t)$ into a mean and a dynamic contribution.
\begin{equation*}
    \vector{\epsilon}(t) = \left < {\vector{\epsilon}} \right > + \Delta {\vector{\epsilon}}(t) ,
\end{equation*}
with $\left < {\vector{\epsilon}} \right > = \frac{1}{T} \int_{0}^{T} \vector{\epsilon}(t) \D t$.
Thus, we can separate the error into a mean and a dynamic term as well.
\begin{equation*}
    \mathcal{E} = \underbrace{\norm{\left < {\vector{\epsilon}}\right >}^2}_{\mathcal{E}_m} +  \underbrace{\frac{1}{T} \int_{0}^{T} \norm{\Delta {\vector{\epsilon}}(t)}^2 \D t }_{\mathcal{E}_d}
\end{equation*}

\begin{figure*}
    \centering
    \subfloat{
        \begin{overpic}{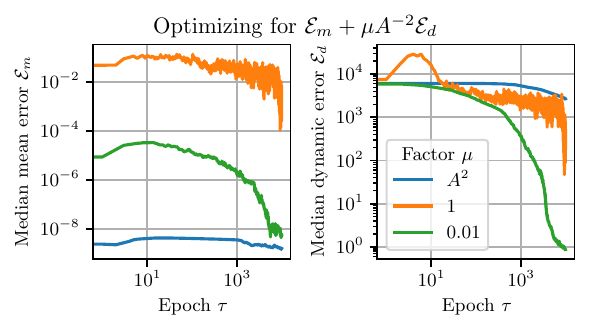}
            \put(15,50){\mybox{A}}
            \put(91,50){\mybox{B}}
            \label{fig:dynamic_weight1}
        \end{overpic}

    }
    \subfloat{
        \label{fig:dynamic_weight2}
    }
    \caption{Median (A) mean and (B) dynamic errors after optimizing for different cost functions. Parameters: (A,B) $\gamma=0.1$, $\omega=0.5$, $A=0.01$, $\phi=0$, $n_s=5$.}
    \label{fig:dynamic_weight}
\end{figure*}

\paragraph{Error scaling}
The dynamic contribution, $\mathcal{E}_d$, scales as $A^2$, while $\mathcal{E}_m$ is independent of the amplitude of the driving.
Rescaling the dynamic contribution, $\mathcal{E}_d$, by $A^{-2}$ allows us to obtain a normalized error that does not depend on the amplitude.
Finally, we can weigh these two terms differently when optimizing the system.
\[
    \begin{array}{l}
        \mathcal E = \mathcal{E}_m + \mathcal{E}_d                               \\
        \mathcal E_\mathrm{norm} \triangleq \mathcal{E}_m + A^{-2} \mathcal{E}_d \\
        \mathcal E_\mathrm{opt} \triangleq \mathcal{E}_m + \mu A^{-2} \mathcal{E}_d
    \end{array}
\]
The weighing factor $\mu$ is a hyperparameter.
We have found that $\mu = 0.01$ gives a good trade-off between the mean error and the dynamic error, as is shown in \cref{fig:dynamic_weight}.

\begin{figure}
    \centering
    \includegraphics{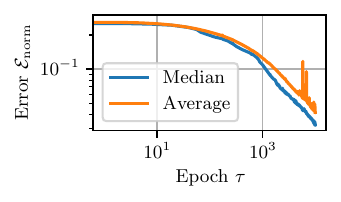}
    \caption{Median and average normalized error over 100 independent realizations. $\gamma=0.1$, $\omega=0.1$, $A=0.1$, $\phi=0$, $n_s=1$.}
    \label{fig:median_vs_average}
\end{figure}

\paragraph{Median vs.\ average}
We use the median rather than the average since it is less susceptible to outliers as shown in \cref{fig:median_vs_average}.

\paragraph{Linear method}
For the linear method, we optimize for the dynamic error obtained from the linearized system and the mean error obtained by using implicit differentiation.
To this end, we need to compute the gradient, $\frac{\partial \vector{r}_t^*} {\partial l_{j0}}$.

First, the system is relaxed using the \acrshort{fire} algorithm~\citep{FIRE}.
This algorithm solves for force balance by adjusting the locations of the nodes $\vector{r}_i$.
Note that the potential energy is a function of the rest lengths $l_{j0}$.
Therefore, in equilibrium, the forces on the nodes vanish:
\begin{equation}
    \vector{F}(\vector{r}_i^*, l_{j0}) = - \vector{\nabla} V(\vector{r}_i^*, l_{j0}) = 0.
    \label{eq:force_balance}
\end{equation}
Here, the star denotes the equilibrium position. \Cref{eq:force_balance} implicitly defines the equilibrium positions as a function of the rest lengths.
\begin{equation*}
    \vector{r}_i^* = \vector{r}_i^*(l_{j0}).
\end{equation*}
Accordingly, as the mean error $\mathcal{E}_m$ is a function of the mean target trajectory $\vector{r}_t^*$, we can write
\begin{equation*}
    \mathcal{E}_m = \mathcal{E}_m(\vector{r}_t^*(l_{j0})).
\end{equation*}
Deriving with respect to $l_{j0}$ and using the chain rule, we get
\[
    \frac{\partial \mathcal{E}_m}{\partial l_{j0}} = \frac{\partial \mathcal{E}_m}{\partial \vector{r}_t^*} \cdot \frac{\partial \vector{r}_t^*} {\partial l_{j0}}.
\]
$\frac{\partial \mathcal{E}_m}{\partial \vector{r}_t^*}$ is trivial to calculate.
$\frac{\partial \vector{r}_t^*} {\partial l_{j0}}$ can be obtained by using the implicit differentiation on \cref{eq:force_balance}.
\begin{align*}
    \frac{\partial \vector F (\vector{r}_i^*, l_{j0})}{\partial l_{j0}} = \partial_{\vector{r}_i^*} \vector F (\vector{r}_i^*, l_{j0}) \cdot \frac{\partial \vector{r}_i^*}{\partial l_{j0}} + \partial_{l_{j0}} \vector F (\vector{r}_i^*, l_{j0}) = 0 \\
    \Rightarrow \frac{\partial \vector{r}_i^*} {\partial l_{j0}} = - \left( \partial_{\vector{r}_i^*} \vector F (\vector{r}_i^*, l_{j0}) \right)^{-1} \partial_{l_{j0}} \vector F (\vector{r}_i^*, l_{j0})
\end{align*}

The dynamic error $\mathcal{E}_d$ (normalized by $A^{-2}$) can be calculated from the steady-state response of the linearized system.
\begin{equation*}
    A^{-2} \mathcal{E}_d = \frac{1}{2} \norm{
        \begin{pmatrix}
            G_x (l_{j0}) \\ G_y (l_{j0})
        \end{pmatrix} - \begin{pmatrix}
                            G_{x,\mathrm{desired}} \\ G_{y,\mathrm{desired}}
        \end{pmatrix}
    }^2 ,
\end{equation*}
where $G_x, G_y \in \mathbb{C}$ are the steady state gain of the linearized system in the $x$ and $y$ directions at the driving frequency $\omega$, i.e.\
\[
    G_x = \frac{\mathcal{F}\{x_\mathrm{target}\}(\J \omega)}{\mathcal{F}\{x_\mathrm{source}\}(\J \omega)}, \quad G_y = \frac{\mathcal{F}\{y_\mathrm{target}\}(\J \omega)}{\mathcal{F}\{x_\mathrm{source}\}(\J \omega)},
\]
where $\mathcal{F}$ denotes the Fourier transform, $x_\mathrm{target}$ and $y_\mathrm{target}$ are the $x$ and $y$ components of the target trajectory, and $x_\mathrm{source}$ is the $x$ component of the source trajectory (assuming that the source only moves in the $x$ direction).
For motion in the $x$-direction, where the target obtains equal amplitude as the source, we set $G_{x,\mathrm{desired}} = \exp (\J \phi)$ and $G_{y,\mathrm{desired}} = 0$, where $\phi$ is the phase difference between input and output.
$G_x$ and $G_y$ are explicit functions of $l_0$ and therefore backpropagation can be used to differentiate the dynamic error.

\paragraph{Optimization}
The gradients are calculated by using JAX-MD~\citep{jaxmd}, which is powered by JAX~\citep{jax}.
In addition, JAXopt~\citep{jaxopt} is used to calculate the gradient of the mean error for the linearized system by using the implicit differentiation.
We use gradient descent to adapt the rest lengths $l_0$.
\begin{equation*}
    l_0 \leftarrow l_0 - \eta \nabla_{l_0} E_\mathrm{opt}
\end{equation*}
The adaptation of the rest lengths is limited to $\pm 50 \%$ of the original rest lengths of the springs.

\subsection{Two-step relaxation: nonlinear method}
\label{appendix:two_step_relaxation_nonlinear}
We have found that the nonlinear method diverges from the linear method at a given epoch during optimization.
With Fourier analysis, we can separate the error into a linear and a nonlinear part.
The linear part contains the mean error (DC contribution) and the linear part of dynamic error ($\pm\omega$ contribution).
The nonlinear part is taken to be the remaining harmonics.

\subsection{Two-step relaxation: linear method}
\label{appendix:two_step_relaxation_linear}
The error in the linear method for low $\gamma$ and $\phi = 0$ also exhibits two-step relaxation.
The dynamic error can be separated into its in-phase and $\frac{\pi}{2}$ out-of-phase contributions.
For example, say that we desire the target to move according to $A \sin (\omega t)$, while its actual motion is given by $B \sin (\omega t) + C \cos (\omega t)$.
Therefore, $|A - B|^2$ is an indication of the in-phase error, while $|C|^2$ is an indication of the 90\textdegree{} out-of-phase error.
\[
    \mathcal{E}_d = \mathcal{E}_{d, 0} + \mathcal{E}_{d, \frac{\pi}{2}}
\]
These two terms of the dynamic error don't necessarily decrease at the same rate, which may lead to two-step relaxation.

\subsection{Lyapunov exponent}
\label{appendix:lyapunov}
\begin{figure}
    \centering
    \includegraphics{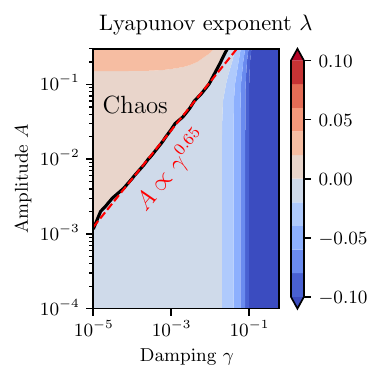}
    \caption{The Lyapunov exponent $\lambda$ as a function of $\gamma$ and $A$. $\omega=0.5$.}
    \label{fig:lyapunov_phase_plot}
\end{figure}

To assess the Lyapunov exponent we integrate the equations of motion (without optimization) for two copies of the system, with slightly different initial conditions.
The difference is in the initial positions of the nodes, which are randomly displaced with a zero-mean normal distribution, and a standard deviation of $\sigma = A / 10$.
The difference between the position of the $i$-th node in both simulations is given by $\delta \vector{r}_i (t)$.
We then average this difference over all the nodes.
\begin{equation*}
    \delta Z^{(m)} (t) = \frac{1}{M^{(m)}} \sum_{i=1}^{M^{(m)}} \| \delta \vector{r}^{(m)}_i (t) \|_2^2
\end{equation*}
with $M^{(m)}$ the number of nodes in realization $m$.
We then take the median over all realizations 100 realizations.
\begin{equation*}
    \delta Z(t) = \mathrm{med}( \delta Z^{(1)} (t), \delta Z^{(2)} (t), \ldots, \delta Z^{(100)} (t))
\end{equation*}
Finally, we can find the Lyapunov exponent $\lambda$ by fitting the following function to the data, making sure that the fit is applied only for short-time scales.
\begin{equation*}
    \delta Z(t) \approx \mathrm \E^{\lambda t} \delta Z(0)
\end{equation*}
The resulting Lyapunov exponent is shown in \cref{fig:lyapunov_phase_plot} as a function of $\gamma$ and $A$.
To create this figure, the spring networks were driven for 100 periods at a frequency $\omega=0.5$.

\subsection{Steady state exponent}
\label{appendix:steady_state}
\begin{figure}
    \centering
    \includegraphics{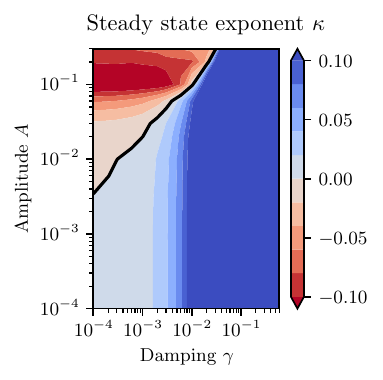}
    \caption{The steady state exponent $\kappa$ as a function of $\gamma$ and $A$. $\omega=0.5$.}
    \label{fig:steady_state_phase_plot}
\end{figure}

We can check if the system reaches a steady state, and the corresponding time scale, by calculating the steady-state exponent $\kappa$.
To this end, we compare the positions of the nodes at time $t$ and at time $t + T$.
We then average this difference over all the nodes, before integrating over an entire cycle.
We do this for every epoch $\tau$.
\begin{equation*}
    \delta S^{(m)} (\tau) = \frac{1}{T} \int_{\tau T}^{(\tau+1) T} \frac{1}{M^{(m)}} \sum_{i=1}^{M^{(m)}} \| \vector{r}^{(m)}_i (t+T) - \vector{r}^{(m)}_i (t) \|_2^2 \D t,
\end{equation*}
with $M^{(m)}$ being the number of nodes in realization $m$.
We then take the median over all 100 realizations.
\begin{equation*}
    \delta S(\tau) = \mathrm{med}( \delta S^{(1)} (\tau), \delta S^{(2)} (\tau), \ldots, \delta S^{(100)} (\tau))
\end{equation*}
The steady-state exponent is found by fitting an exponential function to the data, again making sure that the fit is applied only for short-time scales.
\begin{equation*}
    \delta S(\tau) \approx \mathrm e^ {- \kappa n} \delta S(0)
\end{equation*}
The resulting steady state exponent is shown in \cref{fig:steady_state_phase_plot} as a function of $\gamma$ and $A$.
To create this figure, the spring networks were driven for 100 periods at a frequency $\omega=0.5$.
Comparing this figure to \cref{fig:lyapunov_phase_plot}, we see that the steady-state exponent is negative in the chaotic region, indicating that no steady state is reached.

\section{Supplementary results}
\subsection{Required memory span}
\label{appendix:error_memory_span}
\begin{figure*}
    \centering
    \subfloat{
        \begin{overpic}{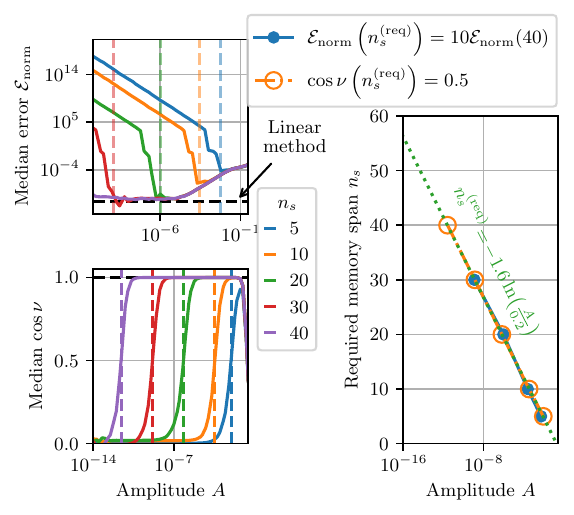}
            \put(7,80){\mybox{A}}
            \put(7,43){\mybox{B}}
            \put(71,15){\mybox{C}}
        \end{overpic}
        \label{fig:error_cosnu_amplitude_fit1}
    }
    \subfloat{
        \label{fig:error_cosnu_amplitude_fit2}
    }
    \subfloat{
        \label{fig:error_cosnu_amplitude_fit3}
    }
    \caption{
        (A) Median normalized error as a function of the amplitude $A$ for different memory spans.
        The vertical lines indicate the departure from the asymptotic curve.
        (B) Average cosine angle between the gradients of the nonlinear and the linear method as a function of $A$ for different memory spans.
        The vertical lines indicate the onset of low overlap, defined as a cosine angle of 0.5.
        (C) Memory spans as a function of the vertical lines in A and B (departure point and low cosine angle).
        Paramters: (A-C) $\gamma=0.1$, $\omega=0.5$, $\phi=0$, $\#\mathrm{epochs} \times \eta = 10^3$.
    }
\end{figure*}

In \cref{fig:error_cosnu_amplitude_fit1}, the normalized error is shown as a function of the amplitude $A$ for different memory spans $n_s$.
The vertical lines indicate the points at which the normalized error diverges from the curve with the highest memory span ($n_s=40$).\footnote{The reason for not taking a larger $n_s$ is the computational cost.}
We choose to identify the divergence point with the criterion $\mathcal{E}_\mathrm{norm}(n_s^{(\mathrm{req}})) = 10 \mathcal{E}_\mathrm{norm}(40)$.
In \cref{fig:error_cosnu_amplitude_fit2}, the cosine angle $\cos \nu$ is shown as a function of the amplitude $A$ and for different memory spans, $n_s$.
The vertical lines mark the value of $A$ where $\cos \nu = 0.5$.
The required memory spans $n_s$, corresponding to these vertical lines, are then plotted as a function of $A$ in \cref{fig:error_cosnu_amplitude_fit3}.
We find good agreement between these two metrics.

\subsection{Mode contribution}
\label{appendix:mode_contribution}
To get a better insight into the role of the participation ratio we look at individual realizations.
The limit of small $\gamma$ is of particular interest since the dynamics dissipate little energy.

\begin{figure*}
    \centering
    \subfloat{
        \begin{overpic}{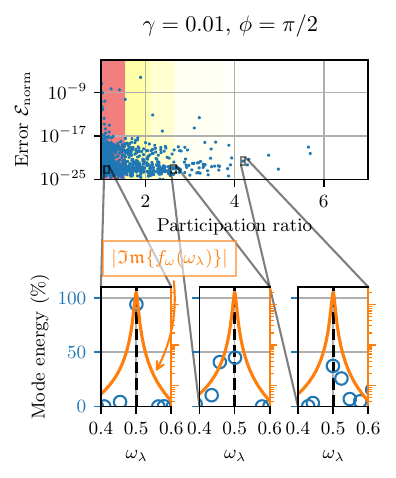}
            \put(21.5,91){\mybox{A}}
            \put(21.5,36.5){\mysmallbox{C}}
            \put(42,36.5){\mysmallbox{D}}
            \put(62.5,36.5){\mysmallbox{E}}
        \end{overpic}
        \label{fig:error_dominant_mode_energy_0.01_1.57}
    }
    \hspace*{-0.5cm}
    \subfloat{
        \begin{overpic}{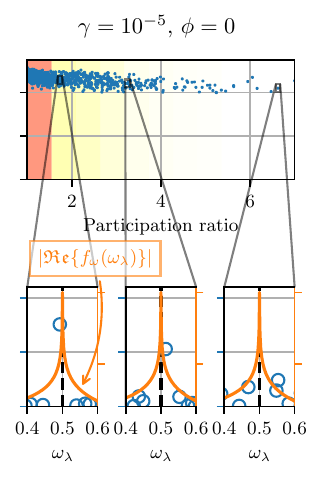}
            \put(5.5,91){\mybox{B}}
            \put(6,36.5){\mysmallbox{F}}
            \put(26.5,36.5){\mysmallbox{G}}
            \put(47,36.5){\mysmallbox{H}}
        \end{overpic}
        \label{fig:error_dominant_mode_energy_1e-05_0}
    }
    \subfloat{
        \label{fig:error_dominant_mode_energy_a1}
    }
    \subfloat{
        \label{fig:error_dominant_mode_energy_a2}
    }
    \subfloat{
        \label{fig:error_dominant_mode_energy_a3}
    }
    \subfloat{
        \label{fig:error_dominant_mode_energy_b1}
    }
    \subfloat{
        \label{fig:error_dominant_mode_energy_b2}
    }
    \subfloat{
        \label{fig:error_dominant_mode_energy_b3}
    }
    \caption{Error vs.\ participation ratio for (A) $\gamma=0.01$, $\phi = \frac{\pi}{2}$ and (B) $\gamma=10^{-5}$, $\phi=0$. The colors in the top figure represent the relative probabilities.
    Mode energy vs.\ eigenfrequency $\omega_\lambda$ for single realizations (C-H). Parameters: (A-H) $\omega=0.5$, $\#\mathrm{epochs} \times \eta = 10^3$.}
\end{figure*}
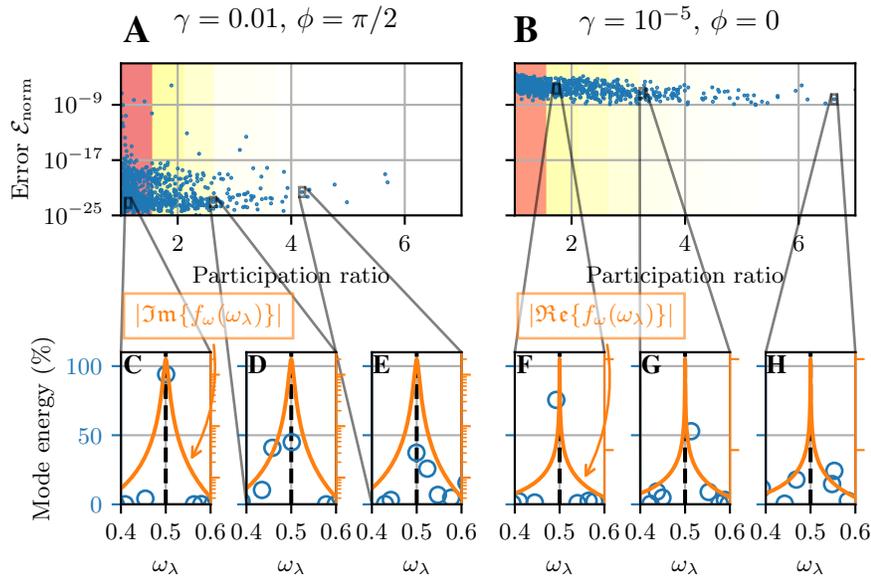

In \cref{fig:error_dominant_mode_energy_0.01_1.57}, we plot the error as a function of the participation ratio for $\phi = \frac{\pi}{2}$.
While most realizations have a low participation ratio and achieve low error, a few outliers with a high participation ratio have an equally low error, meaning that it is not intrinsically necessary to have a low participation ratio to achieve a low error.
The motion can be attributed to a single eigenmode (\cref{fig:error_dominant_mode_energy_a1}), 2 eigenmodes (\cref{fig:error_dominant_mode_energy_a2}), or many eigenmodes (\cref{fig:error_dominant_mode_energy_a3}). Solutions with a single dominant eigenmode around $\omega_\lambda = \omega$ are expected from the shape of the imaginary part of $f_\omega(\omega_\lambda)$.

Similarly, the participation ratio for  $\phi=0$ is shown in \cref{fig:error_dominant_mode_energy_1e-05_0}. The error is slightly larger but the motion is still very close to the desired motion.
Most realizations still make use of a single eigenmode, but the distribution spreads out towards higher participation ratios.
For solutions with a single eigenmode (\cref{fig:error_dominant_mode_energy_b1}), the frequency is situated close to $\omega$.
In contrast to $\phi = \frac{\pi}{2}$, the modes are not arbitrarily close to the driving frequency. This is necessary to suppress the $\frac{\pi}{2}$ out-of-phase motion.
As shown in \cref{fig:error_dominant_mode_energy_b2,fig:error_dominant_mode_energy_b3}, we also observe realizations where the motion of the system is realized using 2 or more modes.

In summary, in the limit of low $\gamma$, we can say that systems prefer to make use of a single eigenmode.
Nevertheless, good results can still be achieved with several eigenmodes.

\subsection{Effect of the learning rate}
\label{appendix:learning_rate}
\begin{figure*}
    \centering
    \subfloat{
        \begin{overpic}{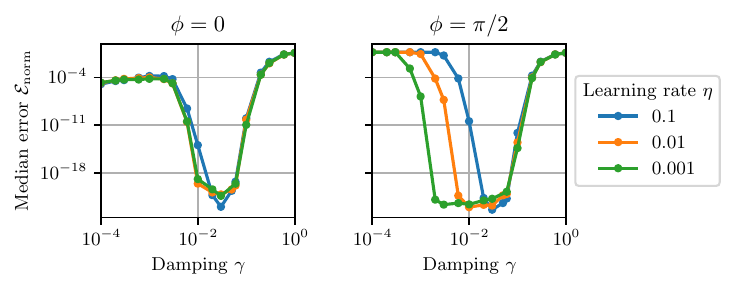}
            \put(15,35){\mybox{A}}
            \put(52,35){\mybox{B}}
            \label{fig:learning_rate1}
        \end{overpic}
    }
    \subfloat{
        \label{fig:learning_rate2}
    }
    \caption{Median (100 realizations) normalized error after training, for (A) $\phi = 0$ and (B) $\phi=\frac{\pi}{2}$, as a function of $\gamma$, for different learning rates $\eta$. Parameters: (A,B) optimized with the nonlinear method with $\# \mathrm{epochs} \times \eta = 10^3$, $\omega=0.5$, $\gamma=0.1$.}
    \label{fig:learning_rate}
\end{figure*}

The learning rate $\eta$ affects the success of optimization differently for $\phi = 0$ and for $\phi = \frac{\pi}{2}$.
For $\phi = 0$ (\cref{fig:learning_rate1}), the learning rate $\eta$ doesn't affect the range of possible damping $\gamma$ that can be attained.
However, for $\phi = \frac{\pi}{2}$ (\cref{fig:learning_rate2}), the learning rate $\eta$ must be lowered progressively as the damping $\gamma$ is decreased.
This is due to the decreasing width of the imaginary part of the weighing function $f_\omega(\omega_\lambda)$ as $\gamma$ decreases.
An eigenmode must be sufficiently close to the driving frequency $\omega$ to get $\frac{\pi}{2}$ out-of-phase motion.
For smaller $\gamma$, the learning rate $\eta$ must be lowered to be able to fit an eigenmode into this small space, without overshooting.

    \bibliographystyle{unsrtnat}
    \bibliography{ref}

\end{document}